\documentclass[twocolumn,amsmath,amssymb,showpacs,prl,superscriptaddress]{revtex4}

\usepackage{epsfig}
\usepackage{color}
\usepackage{graphicx}
\usepackage{bm}
\usepackage{epstopdf}

\begin{document}

\title{Quantum Impurity in a One-dimensional Trapped Bose Gas}

\author{A.~S. \surname{Dehkharghani}}
\affiliation{Department of Physics and Astronomy, Aarhus University,
DK-8000 Aarhus C, Denmark} 
\author{A.~G. \surname{Volosniev}}
\affiliation{Department of Physics and Astronomy, Aarhus University, DK-8000 Aarhus C, Denmark}
\affiliation{Institut f{\"u}r Kernphysik, Technische Universit{\"a}t Darmstadt, 64289 Darmstadt, Germany}
\author{N.~T. \surname{Zinner}}
\affiliation{Department of Physics and Astronomy, Aarhus University,
DK-8000 Aarhus C, Denmark}

\date{\today}

\begin{abstract}
We present a new theoretical framework for describing an impurity in a trapped Bose system in one spatial dimension. The theory handles any external confinement, arbitrary mass ratios, and a weak interaction may be included between 
the Bose particles. To demonstrate our technique, we calculate the ground state energy and properties of a sample system with eight bosons and find an excellent agreement with numerically exact results. Our theory can thus provide definite predictions for experiments in cold atomic gases.
\end{abstract}


\maketitle

\noindent
An impurity interacting with a reservoir of quantum particles is an essential problem of 
fundamental physics. Famous examples include a single charge in a polarizable environment,
the Landau-Pekar polaron \cite{landau1933,pekar1948}, a neutral particle in 
superfluid $^4$He \cite{girardeau1961}, a magnetic impurity in a metal resulting in the Kondo effect
\cite{kondo1964}, and a single scattering potential inside an ideal Fermi gas \cite{fumi1955,kohn1965}. The latter system is famous for the Anderson's orthogonality catastrophe \cite{anderson1967}.
In these settings the 
impurity behavior can provide key insights into the many-body physics and guide our understanding of more general setups.

A complicating feature of many impurity problems is the presence of interactions at a
level that often precludes the use of perturbative analysis and self-consistent mean-field 
approximations. This implies that analytical approaches are highly desirable and exact solutions are, when available, coveted tools for benchmarking other techniques. This is particularly
true for one-dimensional (1D) homogeneous systems where solutions can often be found based on the
Bethe ansatz \cite{bethe1931,lieb1963,mcguire1965,yang1967,lieb1968,guan2013}. These solutions are the essential ingredients for our analytical understanding of highly controllable experiments with cold atoms
\cite{paredes2004,kino2004,kinoshita2006,haller2009,serwane2011,gerhard2012}.
For instance, the exactly solvable problem of the single impurity in a 1D Fermi sea \cite{mcguire1965} - the Fermi polaron - can be used to study the atom-by-atom formation of a 1D Fermi sea \cite{wenz2013}. 

While Fermi polarons have been studied intensively in recent times using cold atomic setups both experimentally and theoretically \cite{schiro2009,nascimbene2009,kohstall2012,koscho2012,massignan2014}, 
the physics of impurities in a bosonic environment is only now becoming a frontier in cold atom 
experiments \cite{catani2012,spethmann2012,scelle2013,fukuhara2013}. This pursuit requires 
theoretical models for describing the Bose polaron \cite{astrak2004,kalas2006,cuc2006,sacha2006,bruderer2008,johnson2012,benja2014,li2014,garcia2014-1,garcia2014-2,amin2014,tempere2009,
peotta2013,rath2013,lych2014,vlie2014,grusdt2014,yulia2014,volosniev2015}, where, in contrast to the Fermi polaron, an exact solution is not known even for a homogeneous 1D system.
Here we provide a new theoretical framework that
captures the properties of an impurity in a bosonic bath confined in one spatial 
dimension.  Our (semi)-analytical theory thus provides 
a state-of-the-art tool for exploring the properties of Bose polarons in 1D.

\begin{figure}[h]
\centering 
\includegraphics[width=\columnwidth]{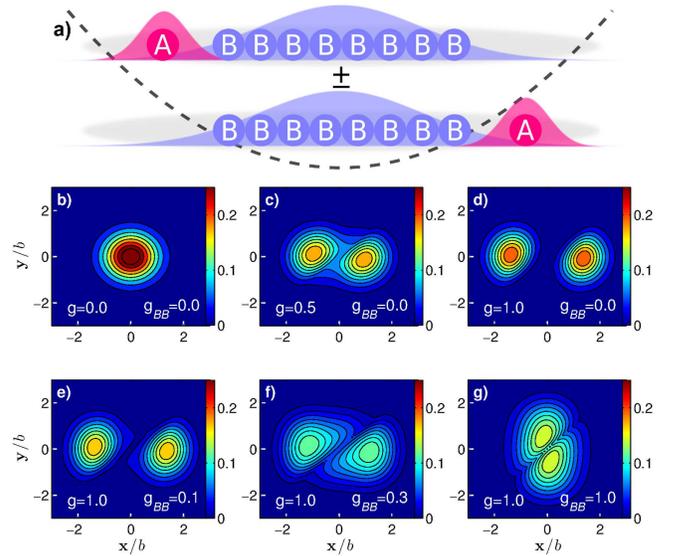}
\caption{ (color online) \textbf{a)} A sketch of the doubly-degenerate ground state for the system of one $A$ and eight $B$ particles trapped in a one-dimensional harmonic potential for infinitely strong repulsive inter-species interaction. The $\pm$ corresponds to the parity of the system. Panels \textbf{b)-d)} show the $AB$ pair-correlation function: the expectation value of the $\delta(z_A-x)\delta(z_B-y)$ operator, where $z_{A(B)}$ is the coordinate of particle $A(B)$. The system under consideration is the same as in \textbf{a)} although with finite inter-species interaction strength,~$g$. Panels \textbf{e)-g)} show the pair-correlation function for different intra-species interaction strength $g_{BB}$ for fixed $g$.}
\label{fig1}
\end{figure}

The proposed framework works with a zero-range potential of any strength
and handles any number of majority particles in external confinement of various shapes which is beyond known
analytical approaches to this problem. Our method is also applicable to describe 
experimental setups that have different trapping potentials for the impurity and majority particles and different mass ratios. In addition, weak majority interactions may be included using the 
well-known Gross-Pitaevskii equation (GPE).  While our 
method is not exact, we have benchmarked the energetics and density 
profiles against numerical results \cite{amin2014} and find agreement for up to ten particles
at the level of a few percent. 
To illustrate the method in this letter, we examine a
system of eight bosons and an impurity in a harmonic trap. Fig.~\ref{fig1}\textbf{a)} shows a sketch of this 
system with vanishing boson-boson and large boson-impurity interaction. This leads to separation of the two components. Notice that the ground
state must retain parity and is thus a linear superposition of the 
two spatial configurations outlined. 
Using the pair-correlation function we can clearly see in Fig.~\ref{fig1}\textbf{b)-d)} how the impurity moves to the 
edge of the system as a function of the inter-species interaction. 
Increasing the intra-species interaction we witness the opposite effect as the impurity goes to 
the center of the system as seen in Fig.~\ref{fig1}\textbf{e)-g)}. 

\paragraph*{Formalism}
Our two component system consists of one type $A$ (impurity), and $N_B$ identical type $B$ bosons (majority) with masses $m_A$ and $m_B$ respectively.  For the sake of argument, we confine particles in harmonic potentials with trapping frequency $\omega_B$ for the bosons and $\omega_A$ for the impurity. In this letter we adopt harmonic oscillator units for the majority particles, i.e. we measure length in units of $b=\sqrt{\hbar/m_B\omega_B}$ and energy in units of $\hbar\omega_B$. Accordingly, the Hamiltonians, for the impurity atom with coordinate $x$ and a majority atom with coordinate $y$, are expressed as
\begin{align}
H_{A}(x)=\frac{p_{x}^2+ m^2_{AB}\omega_{AB}^2 x^2}{2m_{AB}}, \;
H_{B}(y)=\frac{p_{y}^2 + y^2}{2}, \label{eq2}
\end{align}
where $m_{AB}=m_A/m_B$, $\omega_{AB}=\omega_A/\omega_B$,
and $p$ denotes the corresponding momenta.
The interaction between $A$ and $B$ particles is assumed to be of a short range and hence modeled by the Dirac delta-function with strength $g$.
The boson-boson interaction is also given in the standard pseudopotential interaction model \cite{dalfovo} with coupling constant $g_{BB}$. Both interaction strengths are given in units of $[b\hbar\omega_B]$. The overall Hamiltonian of the system is $H=H_{A}(x) + \sum_{i=1}^{N_B} H_{B}(y_i)+\sum_{i=1}^{N_B}g\delta(x-y_i) + \sum_{i<k}g_{BB}\delta(y_i-y_k)$,
where $y_i$ are the coordinates of the bosons (see Supplemental Material \cite{supmat}).

\begin{figure}[t]
\centering
\includegraphics[width=\columnwidth]{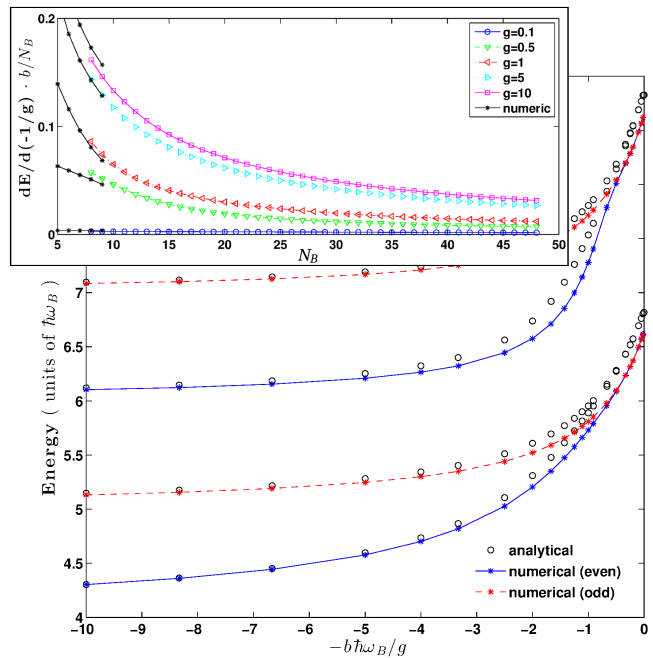}
\caption{(color online) The energy spectrum of low-lying states for a system of eight non-interacting bosons and an impurity with $m_{AB}=\omega_{AB}=1$. The inset shows the derivative of the ground state energy $\partial E/\partial(-1/g)$ divided by $N_B$, which is 
related to Tan's contact parameter. The lines are guides for the eye.}
\label{fig2}
\end{figure}

In order to find the eigenspectrum of the Hamiltonian for arbitrary 
$g$ we introduce a new 
(semi)-analytical approach. 
More specifically, we consider the impurity as the `slow' variable and 
introduce the adiabatic decomposition of the total wave function
\begin{equation}
\Psi(x,y_1,\dots,y_{N_B})=\sum_{j=1} \phi_j(x)\Phi_j(y_1,\dots,y_{N_B}|x),
\label{eq:adiabatic}
\end{equation}
where $\Phi_j$ is the $j$th normalized eigenstate of the eigenvalue problem $\sum_{i=1}^{N_B} H_B(y_i)\Phi_j=E_j(x)\Phi_j$
which we solve assuming that the impurity gives rise to a zero-range potential at a fixed position $x$ (see Supplemental
Material \cite{supmat}).
First we consider an ideal Bose gas, i.e. $g_{BB}=0$, and write $\Phi_j(y_1,\dots,y_{N_B}|x)=\hat{S}\prod_{i=1}^{N_B}f_{k^j_i}(y_i|x)$ with a symmetrization operator $\hat{S}$ (acting on the $y_i$ coordinates) and $f_{k^j_i}(y_i|x)$ being the \textit{$k^j_i$}th normalized eigenstate of $H_B(y_i)$ for a given $x$. 
Notice that every function $f_{k^j_i}(y_i|x)$ has a discontinuous derivative at $y_i=x$ due to the zero-range interaction. This is quantified with the standard delta function boundary condition which dictates that 
the difference in the slopes of the wave function, from the left and right sides of $x$, times a $1/(2g)$ factor must be equal to the value of the wave function taken at $y_i=x$. As $1/g\rightarrow 0$ the wave function must therefore vanish at $y_i=x$.

We can include interactions among the majority particles under the assumption that these may be described by the 1D GPE (see Supplemental Material \cite{supmat}). It has been previously discussed that this is an accurate description for weak interactions between the bosons \cite{ho1999,petrov2000,cominotti2014}.
In this case we need to use a dressed single-particle wave function $\tilde f_{k^j_i}$ instead of $f_{k^j_i}(y_i|x)$. The function $\tilde f_{k^j_i}$ satisfies the 1D GPE complemented with the boundary condition at $y_i=x$,
\begin{align}
\mu(x) \tilde f_{k^j_i}=\left(-\frac{1}{2}\frac{\partial^2}{\partial y_i^2}+\frac{1}{2}y_i^2+N_B\cdot g_{BB}| \tilde f_{k^j_i}|^2\right) \tilde f_{k^j_i}\label{eq3},
\end{align}
where $\mu(x)$ is a chemical potential, $g_{BB}$ is determined through the three-dimensional boson-boson scattering length, $a_s$, as $g_{BB}= \frac{2 a_s}{b}\frac{\sqrt{\omega_{B_1}\omega_{B_2}}}{\omega_B}$, where $\omega_{B_1}$ and $\omega_{B_2}$ are the two frequencies in the directions of strong confinement \cite{bao}. The boson-impurity coupling constant can be also related to the corresponding scattering length \cite{olshanii1998}. 

After determining the function $ f_{k^j_i}$ we obtain a coupled system of equations for $\phi_j(x)$ (see Supplemental Material \cite{supmat}). The coupling terms in this system correspond to the transition of a majority particle from $ f_k$ to $ f_{k'}$. For bosons, this is a coherent process which contributes significantly if the impurity is placed in a region with high density of majority particles. Physical intuition, however, tells us that in the ground and low-lying excited states the impurity is pushed to the edge of the trap if $g_{BB}\ll g$ and $N_B\gg1$.  Otherwise, the impurity would deplete the majority particles notably from the ground state of the one-body harmonic oscillator which is very expensive energywise for $ N_B\gg 1$.  
Hence, for large $N_B$ we neglect the coupling terms between different $\phi_j(x)$, which rigorously gives us an upper bound for the exact energy of the ground state \cite{coelho1991}. However, we expect this approximation to be very accurate also for low-lying excited states.
We can therefore obtain $\phi_j(x)$ by solving numerically a single differential equation. From the mathematical point of view the presented approach is similar to the Born-Oppenheimer approximation or the hyperspherical adiabatic method \cite{nielsen2001}. The physics is, however, different. Indeed, we develop our method for a many-body system where we expect that for the same computational time the relative precision is increasing with the number of particles. 
Clearly, the discussion above applies to a Bose polaron problem in arbitrary confinement. Moreover the trapping potentials as well as the masses can be different for the $A$ and the $B$ particles. Below we show that our framework compares very well to exact numerical results for the equal mass case $m_A=m_B$. In the heavy impurity case with $m_A>m_B$, we expect our model to perform equally well as the impurity becomes increasingly stationary. The 
final case of a light impurity $m_A<m_B$ will not be studied further here, but we note that since it must also go to the edge of the trap, the arguments above for neglegting couplings terms still hold, and we expect our model to work well also in this case.

\begin{figure}[t]
\centering
\includegraphics[width=\columnwidth]{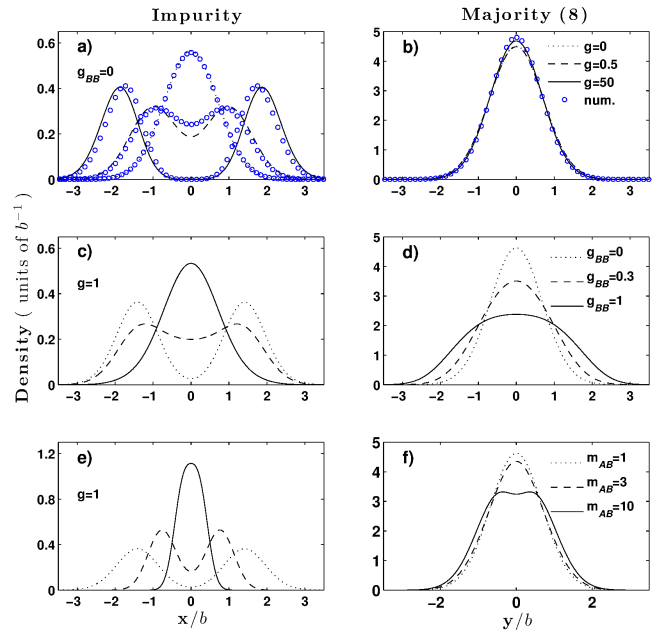}
\caption{ (color online) Panels \textbf{a)-b)} show the density distributions for the impurity, \textbf{a)}, and  majority, \textbf{b)}, atoms for $N_B=8$ and different values of the inter-species interaction strength, $g$. The numerically exact results are shown with blue dots in $\textbf{a)}$ for the three corresponding values of $g$, and in \textbf{b)} for $g=50$. Notice that the label in \textbf{a)} also applies to \textbf{b)} and vice versa. \textbf{c)-d)} The density distributions for a fixed value of $g=1$ but different intra-species interaction strength, $g_{BB}$. \textbf{e)-f)} The density distributions for different values of mass ratios, $m_{AB}=1$, 3, and 10 with $g_{BB}=0$.}
\label{fig3}
\end{figure}

To conclude the presentation of our method we compare its predictions with the exact results obtained using the numerical approach developed in Refs.~\cite{lindgren2014, amin2014} for $g_{BB}=0, m_{AB}=\omega_{AB}=1$.
We find that the relative precision of the method increases with $N_B$ and we pick a sample system with $N_B=8$.
We start by analyzing the energies in Fig.~\ref{fig2}. Our model yields results that are slightly above the numerically exact values with a maximum deviation of a few percent. 
Next we check that the model reproduces the derivative of the ground state energy with respect to the coupling constant, $\partial E/\partial(-1/g)/N_B$, vs. $N_B$, see the inset in Fig.~\ref{fig2}. This derivative for fixed $g$ determines the probability for a given boson to be close to the impurity \cite{barth2011,manuel2012}.  We see that for a large number of bosons this probability becomes smaller, manifesting that the impurity is pushed far from the center of the trap.  It is also interesting to note that for large $N_B$ we find numerically that  $\partial E/\partial(-1/g)$ is almost independent of $N_B$.

We have also compared density profiles and pair-correlation functions for the ground state and again find only minute differences, see Fig.~\ref{fig3}\textbf{a)}~and~\textbf{b)}. 
Note that the impurity density splits for large $g$. 
On the other hand, the majority 
particles are almost unperturbed by the interaction.
This means that an adiabatically slow increase of $g$ moves the impurity to the edge of the 
system, as also shown in Fig.~\ref{fig1}\textbf{a)-d)}. 
As the number of bosons increases the impurity
gets pushed further towards the edge of the trap, and $\partial E/\partial(-1/g)/N_B$ decreases.
In the energy domain it leads to a doubly degenerate ground state at $1/g=0$ since the impurity can be pushed to either the left or the right edges of the trap, see Fig.~\ref{fig2} and Fig.~\ref{fig1}{\bf a)}. 
This should be contrasted with the Fermi polaron system where for large 
interaction the ground state is $N_B+1$ times degenerate and the impurity is localized in the middle of the trap 
\cite{lindgren2014,volosniev2014,levinsen2014}.

\paragraph*{Results}
To further illustrate the model we continue our discussion of an impurity interacting with eight bosons. 
However now we allow weak interaction between the bosons as well as different mass (or frequency) ratios. For large interactions this setup is already beyond current numerical approaches. We stress once again that in our method the numerical complexity does not increase with $N_B$, and $N_B=8$ is chosen as a particular example.
 
First we fix $g=1, m_{AB}=1, \omega_{AB}=1$ and change the intra-species 
interaction strength, $g_{BB}$. In Fig.~\ref{fig3}\textbf{c)} 
and \textbf{d)} we show the density for the impurity and majority atoms in the ground state. In this 
case the energy is minimized if the impurity is pushed towards 
the middle of the trap. This is readily understood for the 
case $g=g_{BB}$ where the impurity particle should have the same 
density distribution due to the boson-impurity exchange symmetry of the Hamiltonian. In our case the difference in the densities of majority and impurity for $g=g_{BB}=1$ is due to the different treatments of these components. Notice also that we expect the GPE and our approach to yield only qualitative results for such a large boson-boson interaction strength. It is also worth noticing that if both interactions $g$ and $g_{BB}$ are very large one can approach the problem directly using the Bose-Fermi mapping \cite{zinner2013}, where one also expects enhancement of the impurity density in the middle of the trap compared to the $g_{BB}=0$ case. Next we compute the pair-correlation function which again demonstrates that the impurity is
situated close to the origin, see Fig.~\ref{fig1}\textbf{e)-g)}.
Consider now different masses for $A$ 
and $B$ particles for $g_{BB}=0, g=1$ and $\omega_{AB}=1$ \cite{omega}. 
%For $m_{AB}\gg 1$ our model is expected to work even better as coupling terms between different $\phi_j$ are weighted by $1/m_{AB}$. 
As shown in Fig.~\ref{fig3} \textbf{e)-f)}, 
when $m_{AB}$ becomes larger the external potential localizes the 
impurity in the middle of the system.  For $m_{AB}\to\infty$ the impurity constitutes a delta-function barrier in the middle of the harmonic trap, the solution to which can be found in Ref.~\cite{busch1998}. From this picture it is apparent that the density of the majority particles should be suppressed at the origin, see Fig.~\ref{fig3}\textbf{f)}.   

\begin{figure}[t]
\centering
\includegraphics[width=\columnwidth]{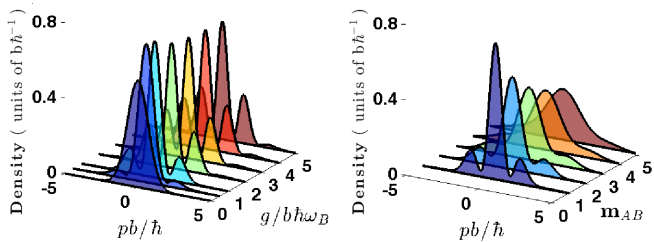}
\caption{Snapshots of the momentum distribution for the impurity 
with different values of $g$ for $m_{AB}=1$ (left) and different values of $m_{AB}$ for $g=1$ (right). We assume $g_{BB}=0$ and $\omega_{AB}=1$.}
\label{fig4}
\end{figure}

Next we consider the momentum distribution which is an
observable to gain information about cold atomic gas systems. 
The 
momentum distributions of the impurity for $g_{BB}=0$ and 
$\omega_{AB}=1$ is shown in Fig.~\ref{fig4} for different mass ratios and 
interaction strengths. These distributions can be understood from the 
discussions above. When the mass ratio increases the 
impurity wave function is almost a Gaussian function, and therefore 
the momentum distribution will also assume a Gaussian form.  Notice the characteristic oscillations in the wings
of the distributions which could be very helpful for the experimental detection of the Bose polaron.
The momentum distribution for majority particles is not plotted because there is no noticeable change in the distribution as we 
change $g$ and/or $m_{AB}$.

\begin{figure}[t]
\centering
\includegraphics[width=\columnwidth]{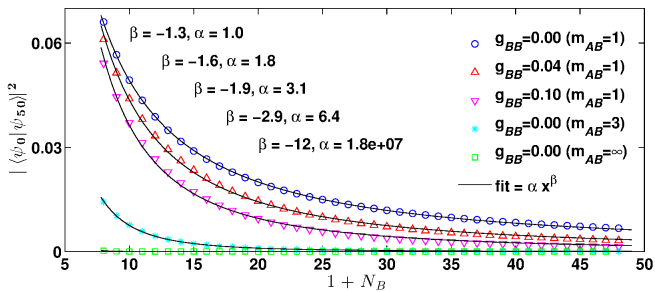}
\caption{(color online) The overlap between the total wave function of the system 
at $g=0$ and $g=50$, $|\langle\psi_{g=0}|\psi_{g=50}\rangle|^2$, 
as a function of $N_B$. The best fit and the corresponding parameters 
are shown in the figure. The upper value of $(\beta,\alpha)$ 
corresponds to the upper curve, etc.
For the infinite mass impurity case, the fit parameters are qualitative 
as there is larger uncertainty here due to the small numerical 
values involved.}
\label{fig5}
\end{figure}

As a final characteristic of the Bose polaron, we consider the overlap 
between the non-interacting and strongly interacting states for different values of $g_{BB}$ and mass 
ratios as function of $N_B$. This quantity is related to the orthogonality 
catastrophe \cite{anderson1967} and has generated recent 
interest as a probe of many-body physics with cold atoms \cite{knap2012,garcia2014-1,levinsen2014}. 
In Fig.~\ref{fig5} we see a power-law behavior, 
but more interestingly, the exponent changes with both $g_{BB}$ and mass ratio.
The overlaps remain finite for finite system sizes and only goes to zero for $N_B\to \infty$ \cite{castella1996}.
The original work of Anderson \cite{anderson1967}
uses a potential to model the impurity which corresponds to the limit $m_{AB}\to\infty$ . 
Consistent with Anderson, this limit shows very fast decay (high negative power dependence on $N_B$)
but already for mass ratio $m_{AB}=3$ the suppression is considerable as seen in 
Fig.~\ref{fig5}.
On the contrary, in the opposite limit of equal masses
we see much longer tails. 
Experiments using equal mass two-component setups and two atomic species
with different masses could therefore complement each other perfectly 
when studying 
the orthogonality catastrophe for Bose polarons.

\paragraph*{Experiments.}
Our predictions should be addressable using current experimental setups. 
In particular, effective 1D systems have been produced that exhibit 
behavior consistent with zero-temperature predictions for both bosonic
\cite{paredes2004,kino2004,kinoshita2006,haller2009} and fermionic atoms \cite{serwane2011,gerhard2012,wenz2013}.
Two-component bosonic systems in 1D \cite{fukuhara2013,fukuhara2013n,hild2014} can be used to explore the equal mass
Bose polarons. Mass-imbalanced Bose-Bose mixtures in 1D have been explored with 
$^{87}$Rb and $^{41}$K ($m_{AB}<1$) \cite{catani2012}
and new experiments with $^{87}$Rb and $^{133}$Cs ($m_{AB}>1$) 
appear promising if an effective 1D geometry can be reached \cite{spethmann2012}. 
Our theory provides predictions for experiments in the 1D regime 
taking into account any experimental features such as different trap frequencies
for different atoms, relative displacement of the trap, and mass imbalance.

\begin{acknowledgments}
We thank M. Valiente and A.~S. Jensen for feedback on the manuscript. The 
authors would also like to thank A.~S. Jensen, D. Fedorov, 
C. Forss{\'e}n, J. Rotureau, and E.~J. Lindgren for collaboration
on strongly interacting 1D systems.
This work was funded by the 
Danish Council for Independent Research DFF Natural Sciences and the 
DFF Sapere Aude program. 
A.G.V. acknowledges partial support by Helmholtz Association under contract HA216/EMMI.

\end{acknowledgments}

\appendix
\widetext
\section{Supplemental Material}
\subsection{The model}

In the main text we consider an impurity particle interacting with $N_B$ bosons. 
There we introduce a decomposition of the total wave function in the following form
\begin{align}
\Psi(x,y_1,\ldots,y_{N_B})=\sum_{j=1} \phi_j(x)\Phi_j(y_1,\ldots,y_{N_B}|x),
\end{align}
where $x$ ($y_i$) is the coordinate of the impurity (majority) atom, and  $\Phi_j$ is a normalized eigenstate of the eigenproblem $\sum_{i=1}^{N_B}H_B(y_i)\Phi_j=E_j(x)\Phi_j$ where we assume that there is a zero-range delta-function interaction sitting at a 
fixed value of $x$. Thus we solve the eigenproblem as a function of the parameter $x$.
Since there are no intra-species interactions among the $B$ particles, we assume a product wave function for the bosons, i.e.
\begin{align}
\Phi_j(y_1,\ldots,y_{N_B}|x)=\hat S\prod_{i=1}^{N_B}f_{k^j_i}(y_i|x), \qquad E_j(x)=\sum_{i=1}^{N_B} \epsilon_{k^j_i}(x),
\label{eq:product}
\end{align}
where $\hat S$ denotes the symmetrization operator acting on the $y_i$ coordinates, and $f_{k^j_i}(y_i|x)$ is the $k_i^j$th normalized eigenstate of the one-boson problem for fixed $x$ with energy $\epsilon_{k^j_i}(x)$. 
Since the position of the impurity is fixed we use free solutions of the corresponding one-body problem on both sides of $x$ and connect those with the delta-function boundary condition, i.e.
\begin{align}
f(x|x)=\frac{1}{2g}\left\{\frac{\partial f(y|x)}{\partial y}\Bigg|_{y=x+\varepsilon}-\frac{\partial f(y|x)}{\partial y}\Bigg|_{y=x-\varepsilon} \right\},
\label{eq:condition}
\end{align}
where the subscript $y=x+\varepsilon$ ($y=x-\varepsilon$) denotes the right (left) derivative at point $x$. For convenience we drop index $k_i$ in $f_{k^j_i}$ and $i$ in $y_i$.
Notice that when $1/g=0$ the boundary condition simplifies to $f(x|x)=0$. As in the main text, we assume a harmonic confinement and adopt natural harmonic oscillator units for a majority particle. Free solutions for the harmonic oscillator potential without requirements of continuity and square-integrability are the Tricomi and Kummer confluent hypergeometric functions multiplied by a Gaussian function \cite{peres2011}. For our problem we need to choose such combinations of these free solutions such that the resulting wave functions are square-integrable and continuous together with their first derivatives everywhere except $y=x$. Moreover at $y=x$ the wave functions should satisfy Eq. (\ref{eq:condition}). For the sake of argument let us consider $x<0$, since the other case is obtained trivially after reflection.  We obtain the following    
 wave function for $y\le x$ 
\begin{align}
f(y|x)&=A(x) e^{-y^2/2} U\left(a(x),\frac{1}{2},y^2\right),
\end{align}
where $a(x)=1/4-\epsilon(x)/2$, $A(x)$ is a normalization factor, $U$ and $M$ are the Tricomi and Kummer confluent hypergeometric functions respectively. The solution for $x<y<0$ reads
\begin{equation}
f(y|x)=A(x) e^{-y^2/2} \cdot \Big( -U\left(a(x),\frac{1}{2},y^2\right) + 2 \frac{U\left(a(x),\frac{1}{2},0\right)}{M\left(a(x),\frac{1}{2},0\right)} M\left(a(x),\frac{1}{2},y^2\right)\Big).
\end{equation}
And on the last interval, i.e. $y>0$, we have
\begin{equation}
f(y|x)=A(x) e^{-y^2/2} \cdot U\left(a(x),\frac{1}{2},y^2\right) \cdot \left( -1 + 2 \frac{U\left(a(x),\frac{1}{2},0\right)M\left(a(x),\frac{1}{2},x^2\right)}{M\left(a(x),\frac{1}{2},0\right)U\left(a(x),\frac{1}{2},x^2\right)}\right).
\end{equation}
The boundary condition from Eq. (\ref{eq:condition}) leads to the following equation for $a(x)$
\begin{align}
\begin{split}
\frac{-2x a(x)}{g} \cdot \frac{U\left(a(x),\frac{1}{2},0\right)}{M\left(a(x),\frac{1}{2},0\right)} \cdot \Bigg( \frac{M\left(a(x),\frac{1}{2},x^2\right)}{U\left(a(x),\frac{1}{2},x^2\right)} &\Bigg. U\left(a(x)+1,\frac{3}{2},x^2\right) + 2 M\left(a(x)+1,\frac{3}{2},x^2\right) \Bigg)+\\
&U\left(a(x),\frac{1}{2},x^2\right)-2\frac{U\left(a(x),\frac{1}{2},0\right)}{M\left(a(x),\frac{1}{2},0\right)}M\left(a(x),\frac{1}{2},x^2\right) =0.
\end{split}
\label{eq:conditiona}
\end{align}
It is interesting to note that if the impurity is placed at the origin, i.e. $x=0$, the above condition can be simplified using the asymptotic values: $\lim_{x\rightarrow0^+}U(a(x),\frac{1}{2},x^2)=\frac{\sqrt{\pi}}{\Gamma(1/2+a(0))}$ and $\lim_{x\rightarrow0^+}U'(a(x),\frac{1}{2},x^2)=\frac{-2a(0)\sqrt{\pi}}{\Gamma(1+a(0))}$. With these values Eq. (\ref{eq:conditiona}) yields the equation for the energy of two interacting atoms in a harmonic oscillator \cite{busch1998}
\begin{align}
(\epsilon(0)-1/2)+g\frac{\Gamma\left(5/4-\epsilon(0)/2\right)}{\Gamma\left(3/2-\epsilon(0)/2\right)}=0 \; .
\end{align}
Once the functions $f(y|x)$ and $\epsilon(x)$ are determined we write down a system of equations for $\phi_j$
\begin{align}
\left[H_A(x)+E_i(x)\right]\phi_i=\frac{1}{m_{AB}}\sum_{j=1} \left(Q_{ij}(x)\phi_j+P_{ij}(x)\frac{\partial \phi_j}{\partial x}\right), \;\; P_{ij}(x)=\langle \Phi_i|\frac{\partial}{\partial x}|\Phi_j\rangle_{y}, \;\; Q_{ij}(x)=\frac{1}{2}\langle\Phi_i|\frac{\partial^2}{\partial x^{2}}|\Phi_j\rangle_{y},
\end{align}
where $m_{AB}$ is the ratio of masses of the impurity and a boson, the subscript $y$ on the brackets denotes integration over all $y_1,\ldots,y_{N_B}$. Since the function $f$ is normalized for every value of $x$ we derive that $P_{ii}=0$ and $Q_{ii}<0$ \cite{nielsen2001}. Let us now estimate the relative contribution of the couplings terms. We start with $P_{jj'}$, this term is not zero if $\Phi_j$ contains only one excitation of bosons compared to $\Phi_{j'}$, i.e. if the corresponding  products in Eq. (\ref{eq:product}) differ by only one function $f$. 
For bosons, excitation is a coherent process. We thus expect $P_{ij}$ to grow with $N_B$ as $\sqrt{N_B}$ for the ground state. 
To understand this in detail we adopt the standard notation of second quantization, i.e. we write $\Phi_j=|n_1,n_2,...,n_k\rangle$, where $n_1,...,n_k$ are occupation numbers in $1,...,k$ states. Without loss of generality let us assume that $\Phi_{j'}=|n_1+1,n_2-1,...,n_k\rangle$. The coupling is then 
\begin{align}
P_{jj'}=\langle n_1,n_2,...,n_k|a_2^\dagger a_1 |n_1+1,n_2-1,...,n_k\rangle \int f_2(y|x)\frac{\partial}{\partial x} f_1(y|x) \mathrm{d}y,
\end{align}
where $a^\dagger$ is the creation operator. This means that $P_{jj'}\sim \sqrt{n_2(n_1+1)}$, so we see that it is smallest if $n_1=0,n_2=1$ and largest if both $n_1\sim n_2\sim N_B/2$. Similarly $Q_{ij}$ is non-zero between states with either zero or one or two relative excitations. Obviously $Q_{ii}\sim N_B$, whereas if the excitation is only one then in analogy with $P_{ij}$ we have $Q_{ij}\sim\sqrt{n_2(n_1+1)}$, where $n_1, n_2$ have the same meaning as above. For two excitations let us consider only the following states $\langle n_1,n_2,...,n_k|$ and $\langle n_1+1,n_2+1,n_3-1,n_4-1,...,n_k|$ for which we obtain that $Q_{ij}\sim\sqrt{n_3 n_4 (n_1+1)(n_2+1)}$. Hence we conclude that only terms $Q_{ij}$ that correspond to none or two excitations should be kept for large $N_B$ if one considers the ground or low-lying excited states. From physical intuition we expect the coupling terms to contribute only if the impurity is in the middle of the trap, which corresponds to large values of $\frac{\partial f}{\partial x}$. At the same time, we expect that for strong interactions the impurity is pushed to the side of the trap to minimize the total energy of the system in the ground state and in low-lying excited states. This allows us to neglect off-diagonal coupling terms $Q_{ij}$. Let us for clarity consider the ground state, 
then we find the impurity wave function, $\phi_1(x)$, numerically from the equation
\begin{equation}
\begin{split}
&\Bigg(H_A(x)+N_B\epsilon_{gs}(x)+\frac{N_B}{m_{AB}}\left\langle\left(\frac{\partial f_{gs}(y|x)}{\partial x}\right)^2\right\rangle_y\Bigg) \phi_1(x)=E\phi_1(x)
\label{numericalequation}
\end{split}
\end{equation}
where $f_{gs}(y|x)$ is the ground state wave function of the one-boson problem for fixed $x$ with energy $\epsilon_{gs}(x)$. Notice that the energy $E$ provides a variational upper bound to the exact energy \cite{coelho1991}.

\subsection{The density and pair-correlation function.} In our analytical results we use the $N$-body wave function, $\psi(x,y_1,\dots,y_{N_B})$  to obtain the densities of impurity and majority components,
\begin{equation}
n(x)=\int{|\psi|^2\mathrm{d}y_{1}\mathrm{d}y_{2}\dots \mathrm{d}y_{N_B}}, \;\qquad n_B(y)=N_B\int{|\psi|^2\mathrm{d}x\mathrm{d}y_{2}\mathrm{d}y_{3}\dots \mathrm{d}y_{N_B}},
\end{equation}
and the pair-correlation function for a boson-impurity pair,
\begin{equation}
n_{AB}(x,y_1)=\int{|\psi|^2 dy_{2}\dots dy_{N_B}}.
\end{equation}
To find the momentum distribution we first define the Fourier transform
of the wavefunction:
\begin{equation}
\psi(p, q_1,\dots,q_{N_B})=\left(\frac{1}{\sqrt{2\pi}}\right)^{1+N_B}\int_{all~space}{\psi(x_1,y_1,\dots,y_{N_B})e^{ipx}e^{iq_1y_1}\dots e^{iq_{N_B}y_{N_B}}}dx_1dy_1\dots dy_{N_B},
\end{equation}
where $p$ ($q_i$) is the momentum of the impurity (majority) particle. 
Next we obtain the momentum distribution for the impurity using the assumption that the wave function is factorized,
\begin{equation}
n(p)=\frac{1}{2\pi}\int_{all~space}{\phi^*(x)\phi(\tilde{x})\left(\int f^*(y|x)f(y|\tilde{x}) \right)^{N_B}~e^{ip(x-\tilde{x})}~dxd\tilde{x}}.
\end{equation}\\

\subsection{The Gross-Pitaevskii equation in a quasi-one dimensional geometry}
For convenience of the reader we review the Gross-Pitaevskii equation (GPE) which describes the ground state for a system of weakly-interacting identical bosons at zero temperature that is specified with the following three-dimensional Hamiltonian with $\hbar=1$
\begin{equation}
H=\sum_{i=1}^{N_B} \left(-{\frac{1}{2m_B}}{\nabla^2}+V({\textbf{r}_i})\right)
+\sum_{i<j}G_{BB}\delta({\textbf{r}}_i-{\textbf{r}}_j),
\end{equation}
where  ${\textbf{r}}_i$ is the coordinate of the $i$-th boson, $m_B$ is the mass of a boson. For small values of the boson-boson scattering length $a_s$ the coupling constant reads $G_{BB}=\frac{4\pi a_s}{m_B}$, see e.g. \cite{abrikosov}.  The external potential is taken to be a harmonic oscillator, $V({\textbf{r}_i})=\frac{1}{2} m (\omega_x^2 x^2+\omega_y^2 y^2+\omega_z^2 z^2)$. Finally, $\delta({\textbf{r}})$ is the 3D Dirac delta-function.
This Hamiltonian describes a weakly-interacting system meaning that the $s$-wave scattering length is much smaller than the average distance between atoms.
To obtain the GPE we assume that the total wave-function $\Psi$ of the system of $N_B$ bosons is a product of single-particle functions $\psi$,
\begin{equation}
\Psi({\textbf{r}}_1,{\textbf{r}}_2,\dots,{\textbf{r}_{N_B}})=\psi({\textbf{r}}_1)\psi({\textbf{r}}_2)\dots\psi({\textbf{r}_{N_B}}),
\end{equation}
which minimizes the expectation value of the Hamiltonian. To satisfy this requirement the single-particle wave-function $\psi(\textbf{r},t)$, should solve the GPE,
\begin{equation}
\mu \psi=\left(-\frac{1}{2m_B}{\nabla^2}+V({\textbf{r}_i}) + G_{BB}N_B\vert\psi\vert^2\right)\psi,
\end{equation}
where $\mu$ is the chemical potential. The chemical potential appears in the equation as a Lagrange multiplier from the normalization condition for the wave function, i.e. $\int dV |\psi|^2=1$. 
To simulate one-dimensional geometry experimentally one confines a system by a potential with two frequencies that are much larger than the third. At very low temperature this means that any two atoms outside of the interaction range move effectively in 1D. Having this in mind we scale the GPE using the units given by the frequency of weak confinement: $\textbf{r}\rightarrow\textbf{r}b$ and $\psi(\textbf{r},t)\rightarrow\psi(\textbf{r},t) b^{-3/2}$, where $b=\sqrt{\frac{1}{m_B\omega_y}}$, such that
\begin{equation}
\frac{\mu}{\omega_y} \psi=\left(-\frac{1}{2}{\nabla^2}+V(\textbf{r})+ \frac{4\pi a_s N_B}{b}\vert\psi\vert^2\right)\psi, \qquad V(\textbf{r})=\frac{1}{2}(\gamma_x^2 x^2+y^2 + \gamma_z^2 z^2)
\end{equation}
where $\gamma_x=\omega_x/\omega_y$ and $\gamma_z=\omega_z/\omega_y$.
Let us now assume a cigar-shaped trap with $\omega_x, \omega_z \gg \omega_y$. This means that the time evolution does not cause excitations along the $x$- and $z$-axis, since these require much more energy than the excitations along the $y$-axis, see e.g. \cite{bao}. We therefore assume that a single-particle solution of the GPE has the following form: $\psi(x,y,z)=\psi_1(y)\psi_{23}(x,z)$, where the function in $x$- and $z$-coordinates is the ground state of the harmonic oscillator: $\psi_{23}=\left(\frac{\gamma_x\gamma_z}{\pi^2}\right)^{1/4}e^{-(\gamma_x x^2+\gamma_z z^2)/2}$. This means that $(x,z)$ part can be easily integrated out which leads to
\begin{equation}
\frac{\mu}{\omega_y}\psi_1=\left(-\frac{1}{2}\frac{\partial^2}{\partial y^2}+\frac{1}{2}y^2+N_B\cdot g_{BB}\vert\psi_1\vert^2\right)\psi_1,
\end{equation}
where $g_{BB}$ is given as:
\begin{align*}
g_{BB}&=\frac{4\pi a_s}{b} \cdot \int{\vert\psi_{23}\vert^4 dxdz}=\frac{4\pi a_s}{b}\frac{\sqrt{\gamma_y\gamma_z}}{2\pi}=\frac{2 a_s \sqrt{\omega_x\omega_z}}{b\omega_y}.
\end{align*}
This equation supplemented with the delta-function boundary condition at the point where a boson meets the impurity is used in the main text to describe weak interactions among the majority particles.


\begin{thebibliography}{99}
\bibitem{omega} Notice that changing the external potentials by tuning the frequency ratio, $\omega_{AB}$, should yield a similar behavior.
\end{thebibliography}
\end{document}